\documentclass[a4paper,11pt]{article}
\usepackage{pos}

\title{The rho-pi puzzle and vector glueball mixing}

\author*[a]{Arthur Vereijken}

\affiliation[a]{Institute of Physics, Jan Kochanowski University, ul. Uniwersytecka 7, 25-406, Kielce, Poland}

\emailAdd{arthur.vereijken@gmail.com}

\abstract{The $\psi(3686)$ is identified as the radial excitation of the $J/\psi$. Based on perturbative QCD, the branching ratio of the $\psi(3686)$ into some final hadron state should be approximately 13\% of the branching ratio of the $J/\psi$ to that same hadron final state. This is called the "13\% rule". However, certain decay channels such as the $\rho\pi$ severely violate this 13\% rule. Using the extended Linear Sigma Model, we study the effect a small mixing angle between the $\psi(3686)$ and the vector glueball can have on the 13\% rule. We show that in a simple model the mixing can already suppress $\psi(3686)$ decays sufficiently to match the $\rho\pi$ observation, but to fully describe the data a more sophisticated model is necessary. We also show that matching to data can tell us about the magnitude of glueball decay widths.}

\FullConference{
}
\begin{document}

\maketitle

\section{Introduction}
\noindent
The $J/\psi$ and $\psi(3686)$ are non-relativistic bound charm-anticharm states. Their primary decay is through 3 gluons, which in the non-relativistic limit is proportional to the wave function at the origin. That is, the decay width to some final state $h$ consisting of light hadrons is \cite{Appelquist:1974zd,Appelquist:1975ya}
\begin{equation}\label{waveft 0}
    \Gamma_{h} = |\psi(\mathbf{r} = 0)|^2 |\mathcal{M}|^2,
\end{equation}
where $\psi(\mathbf{r})$ is the position wavefunction of the system, and $\mathcal{M}$ is the matrix element associated with the decay channel $h$. Up to phase space factors, the decay width is independent of the final state, leading to what is called the $13\%$ rule for the proportion of branching ratios $Q_{h}$ of the $J/\psi$ and the $\psi(3686)$\cite{ParticleDataGroup:2024cfk}:
\begin{equation}
    Q_{h} \equiv \frac{\mathcal{B}(\psi(3686) \to h)}{\mathcal{B}(J/\psi \to h)} \approx \frac{\mathcal{B}(\psi(3686) \to e^{+}e^{-})}{\mathcal{B}(J/\psi \to e^{+}e^{-})} = 13.3\%.
    \label{12 rule}
\end{equation}
Some meson decay channels follow this rule, but in some other channels the rule is severely violated. The $\rho \pi$ channel is the most prominent channel in which this happens. Hence, the name given to this unexplained behavior is "the $\rho \pi$ puzzle".
\\
\\
There are many proposed explanations of the $\rho \pi$ puzzle, see \cite{QuarkoniumWorkingGroup:2004kpm} for a number of examples. One such explanation is that it is caused by mixing with a vector glueball. In the past a mixing between the vector glueball with the $J/\psi$ \cite{Brodsky:1987bb,Hou:1982kh}, as well as the $\psi(3686)$ \cite{Suzuki:2002bz} have been considered.
Given recent lattice predictions for the vector glueball mass in the range of about 3.8 to 4 GeV \cite{Chen:2005mg,Athenodorou:2020ani}, the $\psi(3686)$ is closer in mass to the vector glueball, and so we will assume a mixing between the excited charmonium state and the vector glueball. We will use the framework of the extended Linear Sigma Model (eLSM) \cite{Parganlija:2012fy,Jafarzade:2022uqo}, which has also recently been reviewed in \cite{Giacosa:2024epf}.

\section{Model setup}
\noindent
The central premise of this work is the mixing of the excited charmonium state with a three-gluon $J^{PC}=1^{--}$ vector glueball state. Calling the physical states $\psi(3686)$ as in the PDG listing, and $O'$, the pure charm-anticharm state $\psi_{c\bar{c}}(2S)$, and the pure glueball state $O$, the mixing is as follows:
\begin{equation}
        \begin{pmatrix} 
           \psi(3686) \\
           \mathcal{O}' \end{pmatrix} = \left( {\begin{array}{cc}
   \cos(\theta) & \sin(\theta) \\
   -\sin(\theta) & \cos(\theta) \\
  \end{array} } \right) \begin{pmatrix} 
           \psi_{c\Bar{c}}(2S) \\
           ggg \equiv \mathcal{O} \end{pmatrix}.
           \label{mixing glueball}
\end{equation}
The mixing angle $\theta$ is assumed to be small, so $\psi(3686)$ is predominantly charm-anticharm and $O'$ is primarily gluonic. In order to include phase space corrections to the $13\%$ rule consistent with the eLSM, we introduce the Lagrangian which minimally couples a massive vector field to two fermions:
\begin{equation}
    \mathcal{L}_{V e^{+}e^{-}} = \lambda V_{\mu} \Bar{\psi} \gamma^{\mu} \psi.
    \label{eplus eminus lagranfian}
\end{equation}
Here $\lambda$ is a dimensionless constant which is later divided out, $V_{\mu}$ is the initial massive vector, and $\psi, \Bar{\psi}$ is the (anti)spinor of the electron (positron). The resulting decay width depends on the phase space as
\begin{equation}
    \Gamma_{V e^{+}e^{-}} \propto \frac{|\vec{k}|}{M_{V}^{2}}(M_{V}^2+2m_{e}^2).
    \label{eplus eminus decay}
\end{equation}
With this phase space correction, we can calculate what the 13\% rule \eqref{12 rule} implies for the ratio of the effective couplings $g_{\psi}$ and $g_{J/\psi}$ (essentially the wavefunctions in \eqref{waveft 0}). We find that these couplings are related by
\begin{equation}
    \frac{\cos^2 (\theta) g^2_{\psi}}{g^2_{J/\psi}} \approx 0.35.
    \label{12 rule coupling}
\end{equation}
The $\cos(\theta)$ appears because the glueball component does not directly couple to electrons. The interaction Lagrangians for decays into hadrons are constructed in the eLSM, building on a previous work on decays of the vector glueball in the eLSM \cite{Giacosa:2016hrm}. In this work we will restrict ourselves to the VP channel where the $13\%$ rule is broken most drastically. The decay into PV is described by the Lagrangian
\begin{align}
\mathcal{L}&=  g_{V_{1}}\, \epsilon_{\mu\nu\rho\sigma}\partial^{\rho}V_{1}^{\rho}\text{Tr}[L^{\mu}\Phi R^{\nu} \Phi^{\dagger}] \supset 2\, g_{V_{1}} \,\epsilon_{\mu\nu\rho\sigma}\partial^{\rho}V_{1}^{\rho}\text{Tr}[\partial^{\mu}\mathcal{P}\Phi_{0} V^{\nu} \Phi_{0}],   
\label{VP lagrangian}
\end{align}
where $V_{1}$ is the initial spin vector state, either $\psi_{c\bar{c}}(2S), \,J/\psi, \text{ or } \mathcal{O}$, and $ L^{\mu}:= V^{\mu}+A^{\mu}  \text{, }  R^{\mu}:= V^{\mu}-A^{\mu}$ are left- and right-handed combinations of the vector and axial-vector nonets of light mesons $V^{\mu}$ and $A^{\mu}$. $\mathcal{P}$ is a renormalized pseudoscalar nonet resulting from the shift $A^{
\mu} \to A^{\mu}+\partial^{\mu}\mathcal{P}$ due to spontaneous breaking of chiral symmetry. The vector glueball as well as both the charm anti-charm states are flavor blind with respect to the 3 lightest quarks, meaning they have the same interaction but each with a different coupling indicated by the index in $g_{V_{1}}$. In Eq. \eqref{12 rule coupling} we have a relation between $g_{J/\psi}$ and $g_{\psi}$ that is derived from the $13\%$ rule. The glueball coupling $g_{O}$ and the mixing angle $\theta$ are still unknown constants.

\section{Analysis and results}
\noindent
From the Lagrangian \eqref{VP lagrangian} the decay rate into a vector and a pseudoscalar is given by
\begin{equation}
\Gamma_{V_{1}\to VP} =g_{V_{1}}^2 \kappa_{i} \frac{ |\vec{k}|^3}{12 \pi },
\label{VP gamma}
\end{equation}
where the $\kappa_{i}$ is a dimensionful quantity which depends on the decay products $V \text{ and } P$, it divides out in $Q_{h}$ so the exact values are not important here. The phase space correction here is also different from the one in \eqref{eplus eminus decay}; for the $VP$ channel, the phase space correction will increase $Q_{h}$ compared to the 13\% rule. Due to mixing, the coupling constant for the physical state is transformed as $g_{\psi} \to g_{\psi}\cos(\theta) +  g_{\mathcal{O}} \sin(\theta)$. Using the constraint \eqref{12 rule coupling} and the experimental values of 92.6 KeV and 293 KeV for the total width of the $J/\psi$ and $\psi(3686)$ respectively, $Q_{h}$ becomes a function only of the masses of the particles involved and the combination $\frac{g_{\mathcal{O}}}{g_{J/\psi}}\sin(\theta)$: 
\begin{equation}
    Q_{h} = \left(\frac{g^2_{\mathcal{O}}}{g^2_{J/\psi}}\sin^2(\theta) +2 \sqrt{0.35} \frac{g_{\mathcal{O}}}{g_{J/\psi}}\sin(\theta) + 0.35\right) \frac{ |\vec{k}_{\psi VP}|^3}{ |\vec{k}_{J/\psi VP}|^3} \frac{92.6}{293}.
\end{equation}
We can perform a standard $\chi^2$ fit for the parameter $\frac{g_{\mathcal{O}}}{g_{J/\psi}}\sin(\theta)$ using PDG data. We will restrict ourselves to the $VP$ decay channels that are included in the Lagrangian \eqref{VP lagrangian}, and since at this stage the model treats the charged and uncharged modes identically, we only include the charged $KK^{*}$ mode into the fit. The relevant experimental data and fit results are shown in table \ref{data and results}. The fit yields a reduced $\chi^2$ of $\sim 18$ when mixing with a glueball is included, compared to a reduced $\chi^2$ of about 9600 from only the $13\%$ rule without phase space corrections, and a reduced $\chi^2$ of about 23000 when phase space corrections, but no mixing is included. There are two global minima for $\chi^2$, at  $\frac{g_{\mathcal{O}}}{g_{J/\psi}}\sin(\theta) \approx -0.68$ and at $\frac{g_{\mathcal{O}}}{g_{J/\psi}}\sin(\theta) \approx -0.51$. This degeneracy is due to the relative simplicity of the model, and should vanish when more interactions, for example electromagnetic interactions, are included.
\\
\\
\begin{table}[h]
    \centering
    \begin{tabular}{|c|c|c|c|}
    \hline 
    Decay channel & $Q_{h}$ experimental value \cite{ParticleDataGroup:2024cfk} & Model value (no mixing) & Model value with mixing \\
    \hline
    \hline
    $K^0 \bar{K}^{0*} $ + c.c &  $2.59\pm 0.54 \% $ & 21.1 $\%$ & 0.40 $\%$ \\
         $K^+ K^{-*} $ + c.c &  $0.48\pm 0.10 \% $  & 21.1 $\%$ & 0.40 $\%$\\
            $\eta' \omega $ & $ 16.93\pm 12.28 \%$      & 22.8 $\%$ &  0.43 $\%$ \\
            $\phi \eta $& $4.19\pm 0.54 \%$  & 22.0 $\%$ & 0.41 $\%$ \\
            $ \phi \eta' $ & $3.35 \pm 0.57\%$  & 24.8 $\%$ & 0.47 $\%$ \\
            $\rho \pi$ & $0.19 \pm 0.073 \%$  & 20.0 $\%$ & 0.37 $\%$ \\
            $\eta \rho$ & $ 11.4\pm 3.39 \%$  &  20.8 $\%$ & 0.39 $\%$ \\
            \hline
            \hline
            $\chi^2$ / d.o.f. & -- & 23135 & 18.3 \\
            \hline
    \end{tabular}
    \caption{Data and theoretical results for the ratio $Q_{h}$. Both the model with and without vector glueball mixing have been shown, along with their reduced $\chi^2$. The model with mixing performs significantly better and is in the right order of magnitude for the $\rho\pi$ channel, but does not fully capture the difference between the charged and uncharged channels which seem to be suppressed differently.}
    \label{data and results}
    \end{table}

  \begin{table}
	\centering
		\begin{tabular}[c]{|c|c|}
			\hline
			Decay mode & Decay width (MeV)\\
			\hline \hline
			$\rho \pi$ & 4.65   \\
			\hline
			$K \, K^{*}(892)$ & 6.04\\
				\hline
			$\eta \omega$ & 0.73  \\
			\hline
			$\eta' \omega$ & 0.61  \\
	\hline
			$\eta \phi  $& 0.80  \\
					\hline
     $\eta' \phi $ & 0.69 \\
     \hline
		\end{tabular}
				\caption{Vector glueball decay rates in the PV channel, given a mixing angle of 1\textdegree.}
                \label{tab: glueball decays}
	\end{table}    
\noindent The result of the fit can also be interesting for studying the vector glueball. We can find the coupling $g_{J/\psi}$ from \eqref{VP lagrangian} and \eqref{VP gamma}, through, for example, the $\rho\pi$ channel. Then, choosing\footnote{Which of the two minima we choose does not matter that much as we only roughly estimate the glueballs decays. Choosing the other minimum gives decay widths around 25\% smaller, but the estimate for $\theta$ introduces much more uncertainty.} the minimum $\frac{g_{\mathcal{O}}}{g_{J/\psi}}\sin(\theta) \approx -0.68,$ establishes a relation between $g_{\mathcal{O}}$ and $\theta.$ We can reasonably estimate that the mixing angle is small, but to actually find either of these values, some input from outside is necessary. As a qualitative sketch of what this means, we show the vector glueball decay rates in the $PV$ channel if the mixing angle $\theta$ is 1 degree. Furthermore, for this mixing angle, we estimate the total vector glueball width by comparing to the $J/\psi$ decay width from 3 gluon decays, which is $64\%$ of its total decay width. This estimates the vector glueballs total width at around 90 MeV.
\section{Discussion}
\noindent
The current analysis falls into the camp of "$\psi'$ suppression" hypotheses to explain the ratio $Q_{h}$. We could instead assume a mixing of the vector glueball and the $J/\psi$, i.e. by "$J/\psi$ enhancement". At the current level, this does not change the goodness of the fit. However, the numerics for the place of the minima are different, and so the estimation of the glueball decay width for the same mixing angle is much larger at around 11 GeV, unlikely to be a realistic result.
\\
\\
The estimate on the glueball decay widths has some aspects that need to be mentioned. First of all, the central assumption is that the fit gives an accurate value for the quantity $\frac{g_{\mathcal{O}}}{g_{J/\psi}}\sin(\theta)$, which is based on the premise that the violations of the 13\% rule are (predominantly) due to mixing with the vector glueball, and that the mixing resolves the issue. Looking at the $\chi^2$ shows that this is not fully accurate, so it serves just as a first estimation and proof of concept. Furthermore, the relation $g_{\mathcal{O}}\sim 1/\sin(\theta)\sim 1/\theta$ should not be interpreted physically. From a physical point of view, as $g_{\mathcal{O}}$ increases the glueball couples more strongly to conventional hadrons and so one would expect a larger mixing angle, not smaller. The relation merely states that if the mixing angle is small, the glueball coupling has to be large compared to the $J/\psi$ coupling in order sufficiently suppress $\psi'$ decays.
\\
\\
To recap, using the machinery of the eLSM, and the assumption of a (small) mixing between the vector glueball and the radially excited charm-anticharm bound state, we attempt to explain the $\rho\pi$ puzzle. The resulting fit is a step in the right direction towards agreement with data, especially compared to the unmixed case, but improvements are still necessary to consider the puzzle to be solved. Perhaps the most obvious improvement at this stage is the addition of electromagnetic contributions, in particular the decay through a virtual photon, which according to the PDG accounts for a branching fraction of 1/5th of the 3-gluon decays for the $J/\psi$. In fact, the decay into electron-positron \eqref{eplus eminus lagranfian} is already mediated by a virtual photon. In addition to discriminating between charged and uncharged decay products -- something which is necessary to describe the data but was not done in this work -- it could also cure the degeneracy of there being multiple minima for $\chi^2$.
\\
In its current form, this work can be seen as a step in the right direction towards resolving the $\rho\pi$ puzzle by vector glueball mixing. As evident from table \ref{data and results}, the interference from this mixing can be strong enough to fit to the $\rho \pi$ channel, but more structure is necessary to fully capture the various decay channels. In particular, the decay modes including charged particles are more strongly suppressed than those containing only neutral ones, with the latter being reasonably consistent with a homogeneous suppression to around 3 to 4 percent. In a forthcoming paper we shall expand on this work with a more sophisticated model and analyze more data.

\section*{Acknowledgments}
\noindent
The author is thankful to F. Giacosa for guidance and feedback. This work was supported by the Polish National Science Centre Grant SONATA Project No. 2020/39/D/ST2/02054.

\end{document}